\documentclass{ws-procs11x85}
\usepackage{multirow}
\usepackage{subfigure}

\begin{document}

\title{SEGMENTAL CONVOLUTIONAL NEURAL NETWORKS \\FOR DETECTION OF CARDIAC ABNORMALITY \\WITH NOISY HEART SOUND RECORDINGS\footnote{This work was finished in May 2016, and remains unpublished until December 2016 due to a request from the data provider.}}

\author{YUHAO ZHANG\textsuperscript{1}, SANDEEP AYYAR\textsuperscript{1}, LONG-HUEI CHEN\textsuperscript{2}, ETHAN J. LI\textsuperscript{2,3}}

\address{\textsuperscript{1}Biomedical Informatics Training Program, Stanford University School of Medicine\\
\textsuperscript{2}Department of Computer Science, \textsuperscript{3}Department of Bioengineering, Stanford University\\
Stanford, CA 94305, USA\\
Email: yuhaozhang@stanford.edu, ayyars@stanford.edu, longhuei@stanford.edu, ethanli@stanford.edu}

\begin{abstract}
Heart diseases constitute a global health burden, and the problem is exacerbated by the error-prone nature of listening to and interpreting heart sounds.
This motivates the development of automated classification to screen for abnormal heart sounds. 
Existing machine learning-based systems achieve accurate classification of heart sound recordings but rely on expert features that have not been thoroughly evaluated on noisy recordings.
Here we propose a segmental convolutional neural network architecture that achieves automatic feature learning from noisy heart sound recordings.
Our experiments show that our best model, trained on noisy recording segments acquired with an existing hidden semi-markov model-based approach, attains a classification accuracy of 87.5\% on the 2016 PhysioNet/CinC Challenge dataset, compared to the 84.6\% accuracy of the state-of-the-art statistical classifier trained and evaluated on the same dataset.
Our results indicate the potential of using neural network-based methods to increase the accuracy of automated classification of heart sound recordings for improved screening of heart diseases.

\end{abstract}

\bodymatter
\section{Introduction}
\label{sec1}

Heart diseases constitute a significant global health burden. Just one subset of these diseases, valvular heart disease (VHD) resulting from rheumatic fever, causes 300,000-500,000 preventable deaths each year globally, primarily in developing countries\cite{WHOGeneva2004, Carapetis16122008}. Early detection of many heart diseases is crucial for optimal treatment management to prevent disease progression\cite{Marijon2008, Gersh2015}. In developing countries, the standard practice for screening of heart diseases such as VHD and cardiac arrhythmia is cardiac auscultation to listen for abnormal heart sounds. Patients found to have suspicious abnormalities are then referred to specialists for proper diagnosis by a much more expensive echocardiographic procedure\cite{Marijon2008}. Although cardiac auscultation has been replaced by echocardiography for screening in industrialized countries, the cost-effectiveness and procedural simplicity of auscultation make it an important screening tool for primary care providers and clinicians in under-resourced communities\cite{Sztajzel2010308, Maglogiannis200947}.

The main challenge in cardiac auscultation is the difficulty of detecting and interpreting subtle acoustic features associated with heart sound abnormalities. Manual classification of heart sounds suffers from high intra-observer variability \cite{lok1998accuracy, Ishmail1987870, Jordan1987147, Vukanovic2006, March20051443, Vukanovic2010, Mangione1997, Tavel15031996}, causing false positive and false negative results. Much work has been done in trying to improve screening accuracy, including efforts to design devices to record heart sounds and automatically classify them. However, the biggest challenge for this task remains in developing an accurate classifier for heart sound recordings, which are often obtained in noisy environments. Here, we propose a novel approach based on segmental convolutional neural networks to classification of heart sound recordings. Our approach achieves automatic feature learning together with accurate prediction of the abnormality. On noisy recordings, this approach outperforms prior classifiers using a state-of-the-art feature set developed for noiseless recordings.

The rest of this paper is organized as follows. In Section~\ref{sec2}, we discuss related previous research. In Section~\ref{sec3}, we introduce the methods that we used to classify noisy heart sound recordings, including preprocessing of data, the use of traditional classifiers, and our segmental convolutional neural network models. Next, in Section~\ref{sec4}, we present the performance of our classifiers, along with our analysis of these results. We discuss the limitations of our work and future directions in Section~\ref{sec5} and conclude our work in Section~\ref{sec6}.

\section{Related Work}
\label{sec2}

The first step in automatic classification of heart sounds is segmentation of the recordings along heartbeat cycle boundaries. Segmentation divides the heart sound signal into cycles of four parts: the first heart sound (S1), systole, the second heart sound (S2), and diastole. Past efforts in the field include the use of envelope-based methods\cite{Liang:1997tl, Sun:2014eu} and machine learning techniques\cite{Oskiper:2002gh, Chen:2010va}. A recent segmentation algorithm proposed by Schmidt et al has been shown to work well on a large dataset of 10,172 heart sound recordings, and achieved an average of 95.63\% F1 score, easily outstripping all other methods evaluated with the same set of recordings in the literature\cite{Schmidt:2010cy,Papadaniil:2014ki}. This hidden semi-Markov model (HSMM)-based model was tested on noisy, real-world recordings and considered state-of-the-art. Therefore, we employed the algorithm as-is to acquire the segmentation of input recordings.

Previous work in heart sound recordings classification follows the traditional paradigm of using hand-crafted feature sets as input to automatic classification  based on machine learning. Features are typically a mixture of time domain properties, frequency domain properties, statistical properties, and transform domain properties such as from the discrete wavelet transform (DWT) or empirical mode decomposition (EMD)\cite{Leng:2015iy}. The extracted features are then fed to different machine learning methods, which are then trained to recognize abnormal heart sounds, or in some cases to classify the recordings into the specific heart diseases. The most common methods are artificial neural networks (ANNs)\cite{Uguz:2012fi}, support vector machines (SVMs)\cite{Gharehbaghi:2015ga}, Hidden Markov models (HMM)\cite{Saracoglu:2012wn}, and k-nearest neighbors (kNN)\cite{AvendanoValencia:2010ka}. However, prior results have been restricted by the use of small or otherwise limited data sets, including exclusion of noisy recordings or manual curation of recordings. While classifiers have been reported with accuracies over 90\%\cite{liu2016open}, there is insufficient evidence to conclude whether the expert features used with these classifiers are fully applicable to noisy heart sound recordings. We address this issue by training and testing traditional classifiers on a newly published set of noisy recordings.

In addition, our work is inspired by numerous recent work on the application of neural networks to the processing of sensory-type data, such as visual\cite{krizhevsky2012imagenet, simonyan2014very, wang2012end} and speech recognition\cite{lecun1995convolutional}. However, our segmental convolutional neural network approach is substantially different from these work in its use of heart sound segments during training and test time. Meanwhile, we also empirically evaluated two different types of network architectures and tried to explain their effectiveness via visualizations of learned filters and hidden layers.

\section{Methods}
\label{sec3}

The heard sound recordings used in our experiments were obtained from a publicly hosted dataset\cite{liu2016open} for the 2016 PhysioNet/Computing in Cardiology Challenge\footnote{The data was obtained from the website: https://physionet.org/challenge/2016/}. This dataset consists of approximately 3000 recordings obtained with a variety of durations, noise characteristics, and acoustic features. Cardiac conditions featured in the recordings include valvular heart diseases, benign murmurs, aortic disease, and arrhythmias. Recordings in the dataset were collected from different locations on the body of both children and adults. Given the uncontrolled environment, many recordings are corrupted by various noise sources, such as talking, stethoscope motion, breathing and intestinal sounds, which comprise the challenge of learning features and classifying the signals. As the dataset is intended to support the development of classification systems for initial screening of heart diseases, recordings were only labeled as normal or abnormal, depending on whether follow-up for further diagnosis was recommended from the recording. At the time of our work, recordings excessively corrupted by noise had not yet been relabeled as unclassifiable by challenge organizers, therefore our experiments were focused on binary classification between normal and abnormal noisy recordings, which respectively constituted 80\% and 20\% of the public dataset.

We split the dataset into a 90\% training set for classification model development and a 10\% testing set for model evaluation. In the absence of prior probabilities for disease prevalence, we constructed the test set to be balanced between normal and abnormal recordings for clearer interpretability of performance metrics. As a result, 17\% of the recordings in the training set were abnormal and the remaining 83\% were normal. We preprocessed the recordings and then used them for two independent branches of investigation: traditional classification with feature selection, and the use of a new segmental convolutional neural network architecture. We compared the test set performance of the results of these two investigations by calculating sensitivity, specificity, and accuracy, as these metrics are standard in prior work on heart sound classification. For completeness, we also compared area under the Receiver-Operating Characteristic (ROC) curve and positive predictive value.

\subsection{Preprocessing}\label{sec3.1}

As a first step, we preprocess the recordings to handle noise and segment individual heartbeats. As stated in the related work section, we employ a recent HSMM-based segmentation algorithm developed for noisy heard sound recordings, which has been reported to achieve an accuracy of 95\% on a benchmark dataset.

Since the signals were recorded from multiple sources and differ widely in levels of background noise, we identify the handling of noise within the data crucial for the success of downstream components. We explore a few common venues for denoising in the heart sound recordings and general signal processing, including techniques based on discrete wavelet transform (DWT)\cite{Gradolewski:2014jma} and empirical mode decomposition (EMD). In wavelet-based denoising, the signal is reconstructed from thresholded components produced with DWT using multi-level wavelet coefficients. This approach is finally used in our experiments, given its ease of implementation.

As the noise selection and reduction can vary widely across the multiple data sources and individual cases, we also integrate the denoising results as a feature in traditional classification methods by calculating the signal-to-noise (SNR) of the individual recordings.

\subsection{Traditional Machine Learning-based Classifiers}\label{sec3.2}
We investigated the performance of various machine learning-based classifiers with hand-designed features on noisy heart sound recordings. Through this investigation, we would like to understand: 1) the contribution of different hand-designed features for the classification of noisy heart records; and 2) the overall performance of traditional approaches on this new dataset.

\subsubsection{Features}

We attempt to implement features from a published study for classifying minimal-noise recordings as either normal or abnormal with extraction of 23 features and subsequent selection of 5 features\cite{Singh2013}. Per recording, we extract a set of 58 time-domain, frequency-domain, and transform-domain features which together constituted a superset of the 23 published features, as shown in Table~\ref{table:featuretable}. All features are represented by the mean and standard deviation over all heart beat cycles in the recording. To achieve better results, we transform and combine some features as ratios. Some frequency domain features have missing values due to anomalous recording content.

\begin{table}[ht]
\centering
\tbl{Features extracted for traditional classification.}
{\begin{tabular}{c c c}
\hline\hline
Overall Feature Type & Feature Type & Number of Features \\ [0.5ex]
\hline
\multirow{5}{8 em}{Time Domain} & Interval Length & 16 \\
& Absolute Amplitude & 5 \\
& Total Power & 5 \\
& Zero Crossing Rate & 1 \\
& Amplitude at Peak Frequency & 5 \\
\hline
\multirow{4}{8 em}{Frequency Domain} & Peak Frequency & 5 \\
& Bandwidth & 9 \\
& Q-Factor & 9 \\
& Total Harmonic Distortion & 1 \\
\hline
\multirow{2}{8 em}{Transform Domain} & Cepstrum Peak Amplitude & 1 \\
& Signal-to-Noise Ratio from DWT & 1 \\
\hline
\end{tabular}}\label{table:featuretable}
\end{table}

\subsubsection{Models}

We employ different statistical models to perform supervised learning from the dataset. Before training, we impute all the missing data using median values across all training examples. We perform 10-fold cross validation to evaluate model performance on our 90\% imbalanced training data set. To alleviate classifier bias in learning one class over the other, we follow the standard procedure to increase the weights of classification errors of abnormal recordings. We also use 10-fold cross validation for tuning classifier hyperparameters to improve model performance. Our models and corresponding hyperparameters are shown in Table~\ref{table:hyperparameters}.

\begin{table}
\centering % used for centering table
\tbl{Hyperparameters tuned for the investigated classification models.}
{\begin{tabular}{c c} % centered columns (4 columns)
\hline\hline 
Classification Model & Tuning Parameters \\ [0.5ex]
\hline
Logistic Regression  &-\\
Lasso, Ridge Based Methods &regularizing/penalizing term\\
Support Vector Machine & kernel function, cost, gamma\\
Decision Trees & number of trees\\
Random Forests &number of estimators, number of features\\
K-Nearest Neighbours &number of neighbours\\
\hline
\end{tabular}}\label{table:hyperparameters}
\end{table}
We use forward stepwise, backward stepwise, and Lasso regression methods for feature selection. We use features selected by the forward stepwise and Lasso methods for training the logistic regression classifier and Lasso-selected features for training the remaining classifiers. For the Lasso method, we optimize the regularization/tuning parameter lambda and select features using the lambda value that minimizes the misclassification error rate.

\subsection{Segmental Convolutional Neural Networks for Heart Sound Classification}\label{sec3.3}

Traditional classifiers are simple to employ and fast to train, but rely on hand-designed features that do not necessarily capture useful signals in the recordings. An alternative to traditional classifiers is models that can automatically learn useful features that are not limited by human design. Among these models, convolutional neural network (CNN) provides a flexible filter-based architecture to capture the patterns in the sensory-type data. However, heart sound signals vary in length significantly, and often contain noise that makes a certain snippet of signal unclassifiable. These make the adoption of CNN models less straightforward.

\begin{figure}[h]
\center
\includegraphics[width=0.8\textwidth]{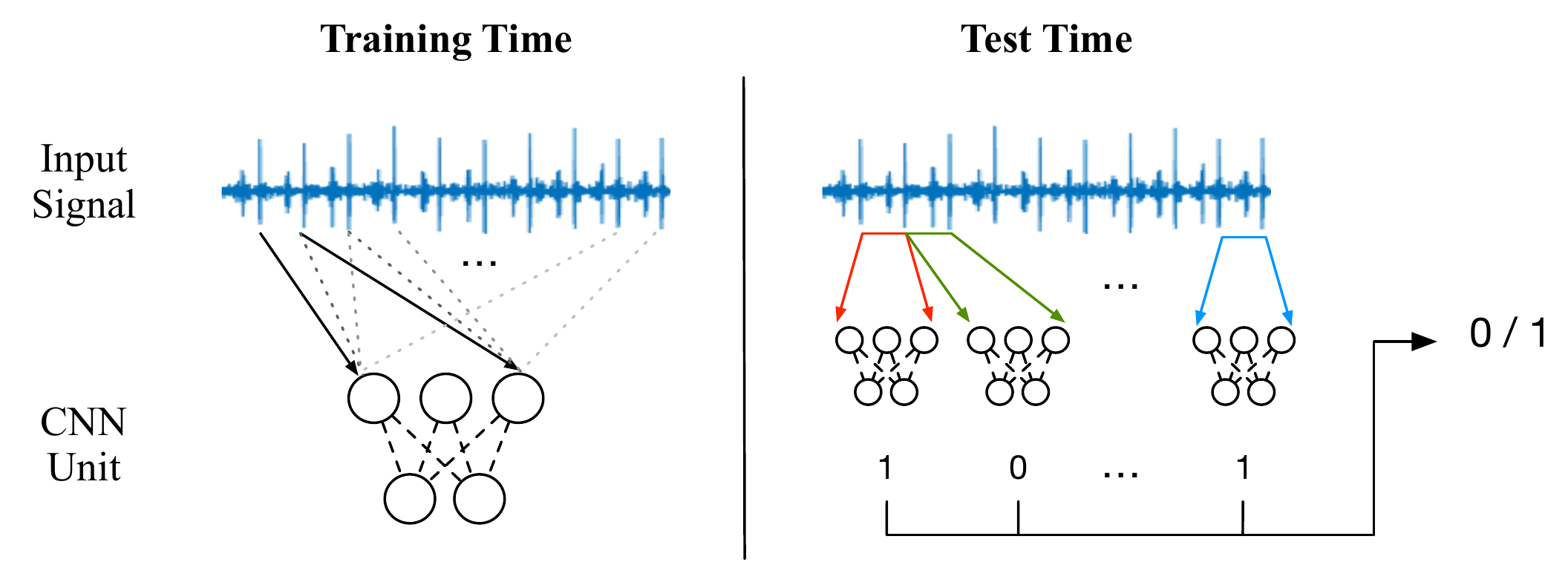}
\caption{Training and evaluation of the segmental convolutional neural networks.}
\label{fig:segmental-cnn}
\end{figure}

We propose a segmental convolutional neural network architecture to solve these problems. As shown in Fig.~\ref{fig:segmental-cnn}, our method takes raw heart sound recordings as input, and acquires recording segments by using the hidden semi-markov model described in Section~\ref{sec3.1}. Then we only keep segments with lengths from 400 to 1200 and zero-pad all signals into a 1200-element vector. During training time, this preprocessing step keeps 98\% of all segments and leaves us with 76509 training segments. We then cast these training segments as a new training set to train our CNN units. During test time, we first split each test signal into segments and then classify each segment using our trained CNN unit. Then we combine the segment classifications and classify a recording as abnormal only when the proportion of segments classified as abnormal is over a threshold. We treat this threshold value as a hyperparameter.

This approach has three key advantages. First, since standard CNN requires fixed-length input, this naturally solves the input length normalization issue. Second, expanding signals into segments substantially increases the number of training instances, which has been proved to be critical in the success of other applications of neural networks. Third, global classification of a recording is more robust against accidental noise in the data, as accidental noise can only influence the classification of a small portion of local segments.

Filter configuration and depth are two major factors that influence the performance of a CNN model. It remains unclear which type of architecture is more suitable to this task. Therefore, we now discuss the use of two different architectures for CNN units, which we name Filter-focused CNNs and Depth-focused CNNs respectively. These two architecture types differ mainly in their configuration of filters, the way max-pooling is conducted, and the way different layers are stacked.

\subsubsection{Filter-focused CNN (FCNN)}

\begin{figure}[h]
\center
\includegraphics[width=0.6\textwidth]{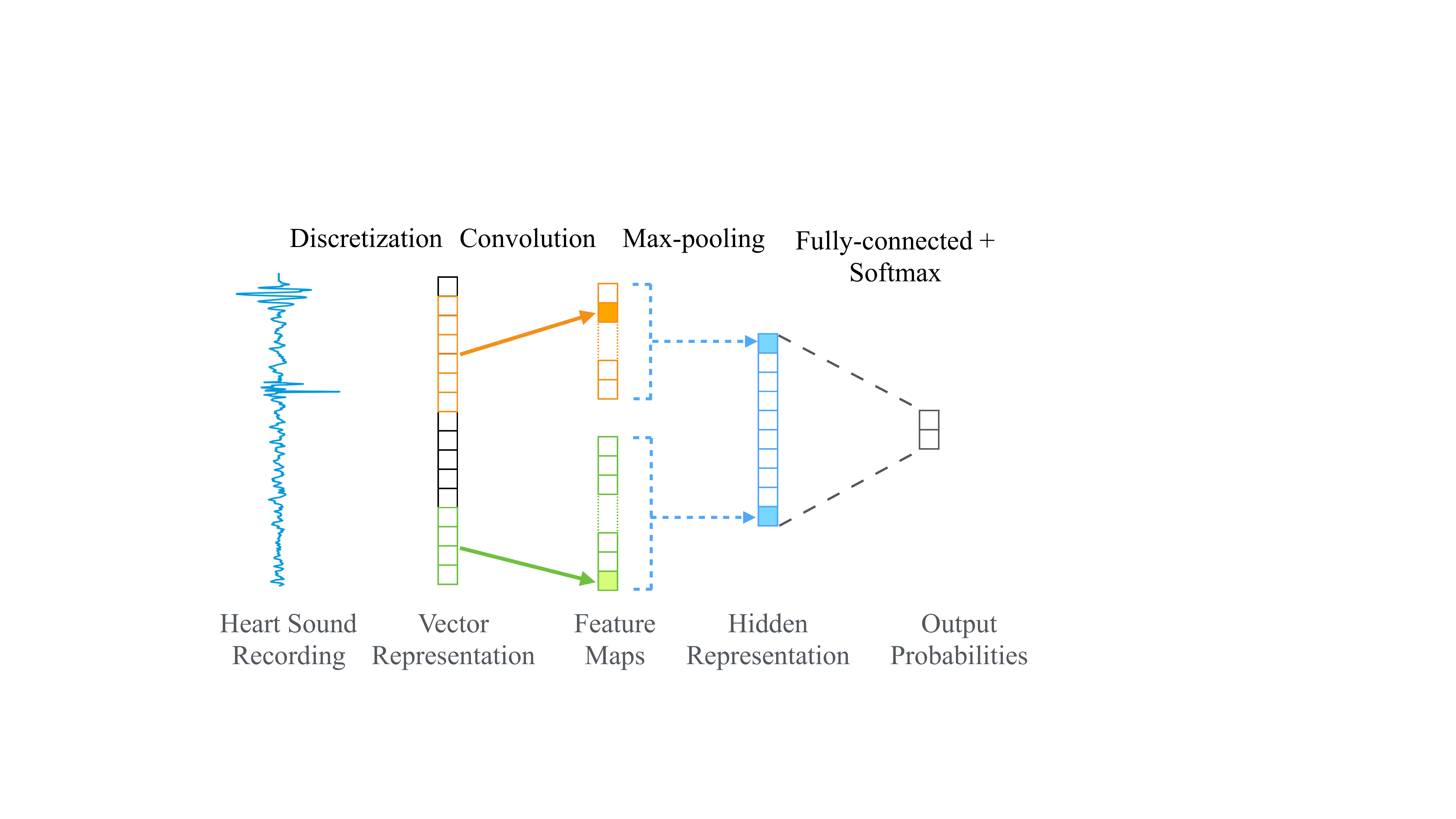}
%\caption{Architecture of a FCNN model. A FCNN model uses many different sizes of filters, and consists of three layers: a convolution layer, a max-over-time pooling layer, and a softmax layer.}
\caption{Architecture of a FCNN model.}
\label{fig:fcnn}
\end{figure}

Fig.~\ref{fig:fcnn} visualizes the architecture of a FCNN model. In a FCNN model, a heart sound segment will first be represented as a vector $x$, with each element in $x$ representing the normalized amplitude of the signal at that time point. The core parameters of the network are a set of filters with different window sizes that will be applied to the input signal $x$. Given a specific window size $w$, a filter is a vector of size $w$ $f = [f_1, f_2, ..., f_w]$, where each $f_i$ is a scalar. A feature map $m$ of this filter can be obtained from the application of a 1D-convolutional operator on $f$ and $x$ to produce an output sequence $m = [m_1, m_2, ..., m_{n-w+1}]$ where $n$ is the length of input signal $x$:
\begin{align*}
    m_i = g(\sum^{w-1}_{j=0}f_{j+1}x_{j+i} + b)
\end{align*}
where $b$ is a bias term and $g$ is a non-linear function. This convolution process is repeated for many filters of different window sizes $w$. After the convolution layer, a max-over-time pooling operation is applied to each feature map $m^{k}$ to generate a single scalar activation $h_k$:
\begin{align*}
    h_k = \max([m^{k}_1, m^{k}_2, ..., m^{k}_{n-w+1}])
\end{align*}
And then all activations $h_k$ are concatenated to form a size-$N$ hidden representation of the original signal $h = [h_1, h_2, ..., h_N]$. The idea behind the max-over-time pooling operation is to only keep the most obvious activation that are generated by the convolution, and use that to characterize the signal for the downstream classification.

Finally, the hidden representation $h$ is fed into a fully-connected with softmax layer to generate the class probabilities $y$. We use cross entropy between predicted labels and ground truth labels as loss function. The intuition behind the use of many filters of various window sizes is that the model should be able to learn through back-propagation common patterns that it has seen in the training signals that are useful for the classification, and these patterns could potentially be numerous and of different scales.

\subsubsection{Depth-focused CNN (DCNN)}

\begin{figure}[h]
\center
\includegraphics[width=0.8\textwidth]{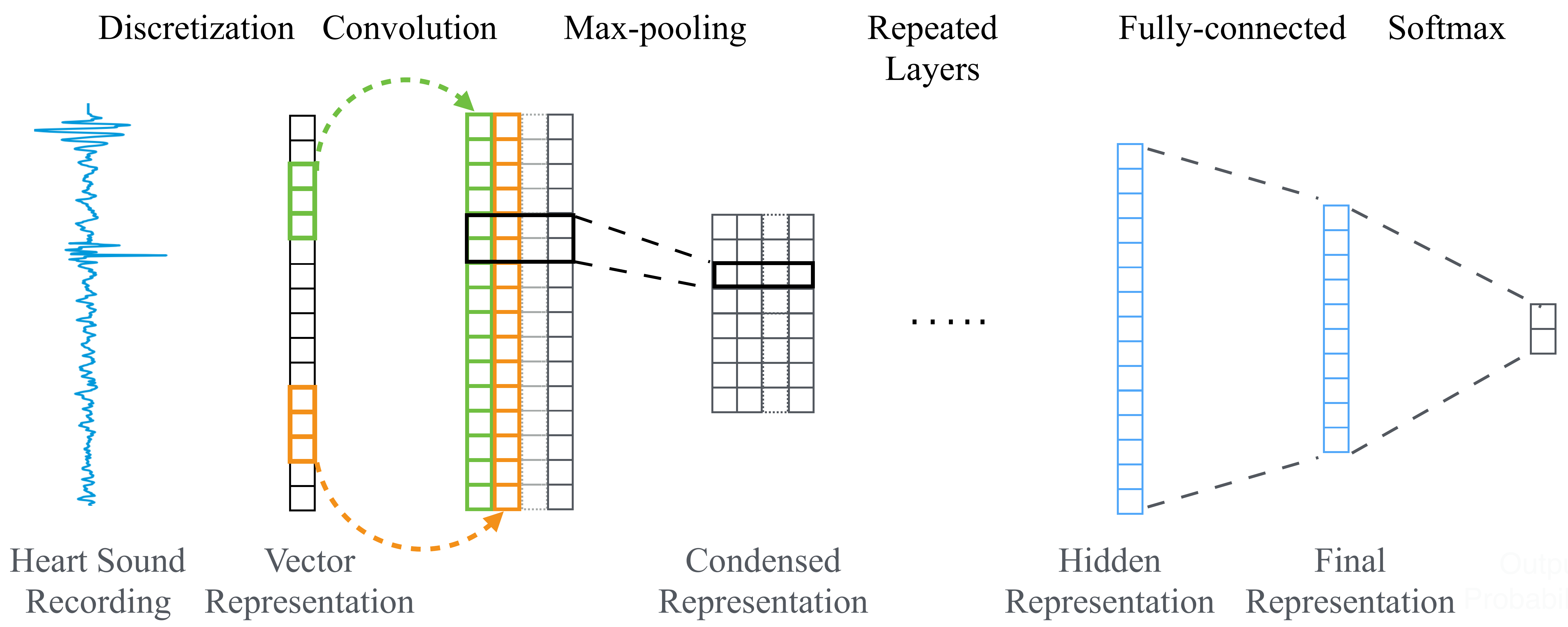}
% \caption{Architecture of a DCNN model. A DCNN model uses a small number of fixed size filters at each convolution layer, and is formed by stacking several convolution-pooling layers, and fully-connected layers together before the final softmax layer.}
\caption{Architecture of a DCNN model.}
\label{fig:dcnn}
\end{figure}

While large filters of various sizes can help to capture useful patterns of different scales, it may also be useful to have a model with only small filters at each layer but focuses on stacking many layers together to form a deep architecture, as has been found in visual recognition tasks \cite{simonyan2014very}. Fig.~\ref{fig:dcnn} visualizes the architecture of a Depth-focused CNN model. There are three major differences between DCNNs and FCNNs. First, the filter sizes in DCNNs are much smaller than in FCNNs. Typically, the size of DCNN filters are approximately 10, while the size of FCNN filters can range from 10 to 500. The use of very small filters in DCNNs reduces the computational cost to perform convolutional operations and thus enables us to explore deeper models while still capturing useful patterns in the signals. Second, in DCNNs, the motif of a convolution layer followed by a max-pooling layer is repeated several times to form a hidden representation of the original signal. Then this hidden representation is fed into multiple stacked fully-connected layers to reduce the representation size, after which the softmax layer generates the output probabilities.

Finally, the way convolution and max pooling are conducted is different. In a DCNN convolution layer, the output of convolution operation is a feature matrix $m = [m_1, m_2, ..., m_n]$, where each column $m_i$ is the feature map vector obtained from filter $f_i$ convolved with signal $x$, and can be viewed as a ``channel'' in the output signal. Then at the pooling layer, instead of doing a max-over-time pooling, a max pooling over the local time region is performed, and channels are kept. For example, a max-pooling operation with window 2 is:
\begin{align*}
    \hat{m}^c_i = \max(m^c_{2i}, m^c_{2i+1})
\end{align*}
where $\hat{m}^c$ represents the pooling output column in channel $c$, and $m^c$ represents the channel $c$ column in the feature matrix. Here, the max pooling serves as a sub-sampling over the signal and preserves more information compared to the max-over-time pooling operation in FCNN.

\subsubsection{Network Configurations}

We design experiments to evaluate our segmental convolutional neural network approach. Table~\ref{tab:cnn-conf} shows the CNN architectures that we report results on. We explored a lot different architectures and included results for these models because: First, these models demonstrate progressively increasing filter sizes and network depths, enabling comparison of the effects of different network configurations on final performance; Second, the training times of these models are tolerable given the resources we have. In the table, ``Conv'' represents a convolution layer, ``MP'' a max-pooling layer, and ``FC'' a fully-connected layer. For instance, for FCNNs, ``Conv([50-500,50]*20)'' represents a convolution layer with window size ranging from 50 to 500 with a step of 50, and each window size corresponds to 20 different filters. For DCNNs, ``Conv([10*25])'' represents a convolution layer with 25 filters and window size 10.

\begin{table}[ht]
\centering % used for centering table
\tbl{CNN network configurations. }
{\begin{tabular}{ |c|c|c|c| } % centered columns (4 columns)
\hline\hline 
CNN Model & Architecture Configuration & \# Layers & \# Filters \\ [0.5ex]
\hline
FCNN-Small &Conv([50-500,50]*20), MP, FC & 3 & 200 \\
FCNN-Medium & Conv([25-500,25]*30), MP, FC & 3 & 600\\
FCNN-Large & Conv([20-600, 20]*50), MP, FC & 3 & 1500\\
\hline
DCNN-Shallow & Conv([10]*25), MP, Conv([10]*50), MP, FC(256), FC
 & 6 & 75\\
DCNN-Deep & Conv([10]*25), MP, Conv([10]*50), MP, Conv([10]*50), MP, FC(256), FC & 8 & 125\\
\hline
\end{tabular}}\label{tab:cnn-conf}
\end{table}

For all CNN configurations, we use L2 regularization on the weights and dropout\cite{srivastava2014dropout} before the last softmax layer to regularize the model. We use AdaGrad\cite{duchi2011adaptive} to train the models with error backpropagation. We train each model on a 90\% subset of our training set for 50 epochs, and after each epoch we evaluate the model on the remaining 10\% validation subset of our training set. For each CNN configuration, we save the model that generates the best accuracy on the validation set as the final model. This allows us to prevent the final model from overfitting on the training data. We then evaluate the best model from each CNN configuration on the same test set as used by the traditional classifiers.

\section{Results} \label{sec4}

\subsection{Traditional Classifiers}
Table~\ref{table:class-result} compares the performance of the traditional classification models. We evaluated model performance based on accuracy, specificity, sensitivity, positive predictive value (PPV), and area under the receiver operating characteristic curve (AUC). The receiver operating characteristic (ROC) curves for all models are shown in Fig. 4. Weighting and feature selection significantly improved performance of most methods except decision trees. Overall, we saw that SVM with feature selection was the best performing model. However, the accuracy of this model on noisy recordings was lower than the published accuracies of models using the same feature set on low-noise recordings, which exceeded 90\% with feature selection\cite{Singh2013}.

Table 5 summarizes the features selected by our models. Contrasting with the previous work which found that four time-domain features and one frequency-domain feature were sufficient for accurate classification of low-noise recordings\cite{Singh2013}, we found that accurate classification of noisy recordings required additional frequency-domain and transform-domain features.

\begin{table}[ht]
\centering % used for centering table
\tbl{Results for different traditional classification models on the test set. The numbers in parentheses show the baseline performances of the models, whereas those outside show performances after model improvement, namely weighting, parameter tuning, and feature selection.}
{\begin{tabular}{ |c|c|c|c|c|c| } % centered columns (6 columns)
\hline\hline 
Classifier & Accuracy (\%) & Specificity (\%) & Sensitivity (\%) & PPV (\%) & AUC (\%)\\ [0.5ex]
\hline
SVM & \textbf{84.6} (69.2) &\textbf{92.2} (68.3) & \textbf{78.3} (56.2) &\textbf{89.2} (63.4) & 83.4 (55.4)\\
Logistic Regression &75.4 (49.5) &74.3 (62.4) &75.1 (50.2) & 74.3 (55.4) & 81.3 (65.4)\\
Random Forests &71.4 (65.4) &91.3 (68.2) &72.3 (65.4) & 81.3 (71.3) & \textbf{88.4} (71.4)\\
KNN (k=3) &60.3 (56.1) &92.2 (65.3) &60.7 (57.6) &  82.4 (79.4) & 71.3 (65.4)\\
Naive Bayes &70.4 (53.4) &60.3 (56.7) &71.5 (53.3) &71.5 (60.6) &77.7 (62.3)\\
Decision Trees &73.3 (73.6) &88.4 (88.6) &73.4 (73.6) &85.5 (85.5)& 69.4 (58.6)\\
\hline
\end{tabular}}\label{table:class-result}
\end{table}

\begin{table}[ht]
\centering % used for centering table
\tbl{Feature selection results.}
{\begin{tabular}{c c}
    \hline\hline
    Feature Type & Number of Features \\ [0.5ex]
    \hline
    Time Domain & 5 \\
    Frequency Domain & 6 \\
    Transform Domain & 2 \\
    \hline
    \end{tabular}}\label{table:featureselection}
\end{table}

\begin{figure}[h]
\center
\includegraphics[width=0.5\textwidth]{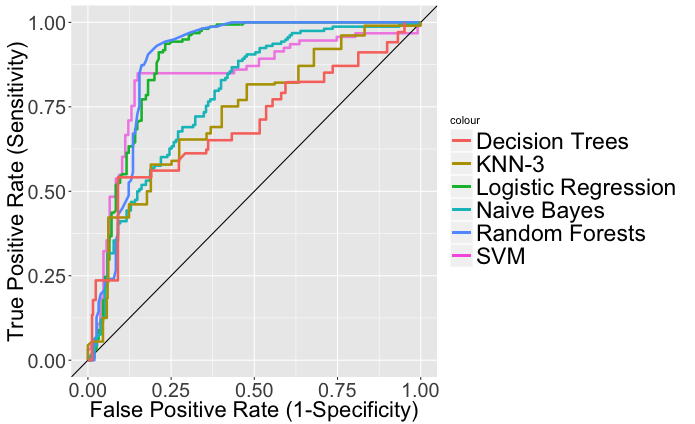}
\caption{ROC curves for classification models.}
\label{fig:ROC}
\end{figure}

\subsection{Segmental Convolutional Neural Networks}

Table~\ref{tab:cnn-result} shows the results of different segmental convolutional neural network models on the test set, with the model names corresponding to configurations in Table~\ref{tab:cnn-conf}. Overall, our DCNN-Deep model produces the best accuracy, specificity and PPV results, while our DCNN-Shallow model produces the best sensitivity. For FCNN models, as filter number increases, we observed an increase in all metrics except sensitivity. This suggests that, in FCNN models, a larger number of filters with more fine-grained window sizes can help the model capture more patterns in the signals, which aligns well with our intuitions. It is worth noting that the FCNN-Large model already produced a relatively high accuracy, and the highest sensitivity value. For DCNN models, as the number of layers increases, we observe increases in almost all metrics (except sensitivity), which suggests that deeper models with more layers can help the model learn better patterns in the signals, which aligns with our assumptions.

In addition, compared to FCNN models, we observed that DCNN models almost always perform better, which suggests that a deep model with small filter sizes and few filters at each layer is more expressive in modeling the heart sound signal data than a shallow model with a large number of filters. However, in terms of sensitivity, we also discovered that the performance does not change much as the filter number and layer number increase. In other words, most of the gain in accuracy comes from the gain in specificity. 

\begin{table}[ht]
\centering % used for centering table
\tbl{Classification results for different CNN configurations, with SVM for comparison.}
{\begin{tabular}{ |c|c|c|c|c| } % centered columns (4 columns)
\hline\hline 
CNN Model & Accuracy (\%) & Specificity (\%) & Sensitivity (\%) & PPV (\%) \\ [0.5ex]
\hline
FCNN-Small & 81.2 & 72.9 & 89.2 & 77.0 \\
FCNN-Medium & 83.4 & 78.1 & 88.6 & 80.5\\
FCNN-Large & 85.3 & 80.0 & \textbf{90.5} & 82.2\\
\hline
DCNN-Shallow & 86.9 & 83.2 & \textbf{90.5} &  84.6\\
DCNN-Deep & \textbf{87.5} & \textbf{88.4} & 86.7 & \textbf{88.4}\\
\hline
SVM & 84.6 & 92.2 & 78.3 & 89.2\\
\hline
\end{tabular}}\label{tab:cnn-result}
\end{table}

\subsection{CNN Visualizations}\label{sec4.3}
% Yuhao

\begin{figure}[h]
\centering
\subfigure[]{\includegraphics[width=0.35\textwidth]{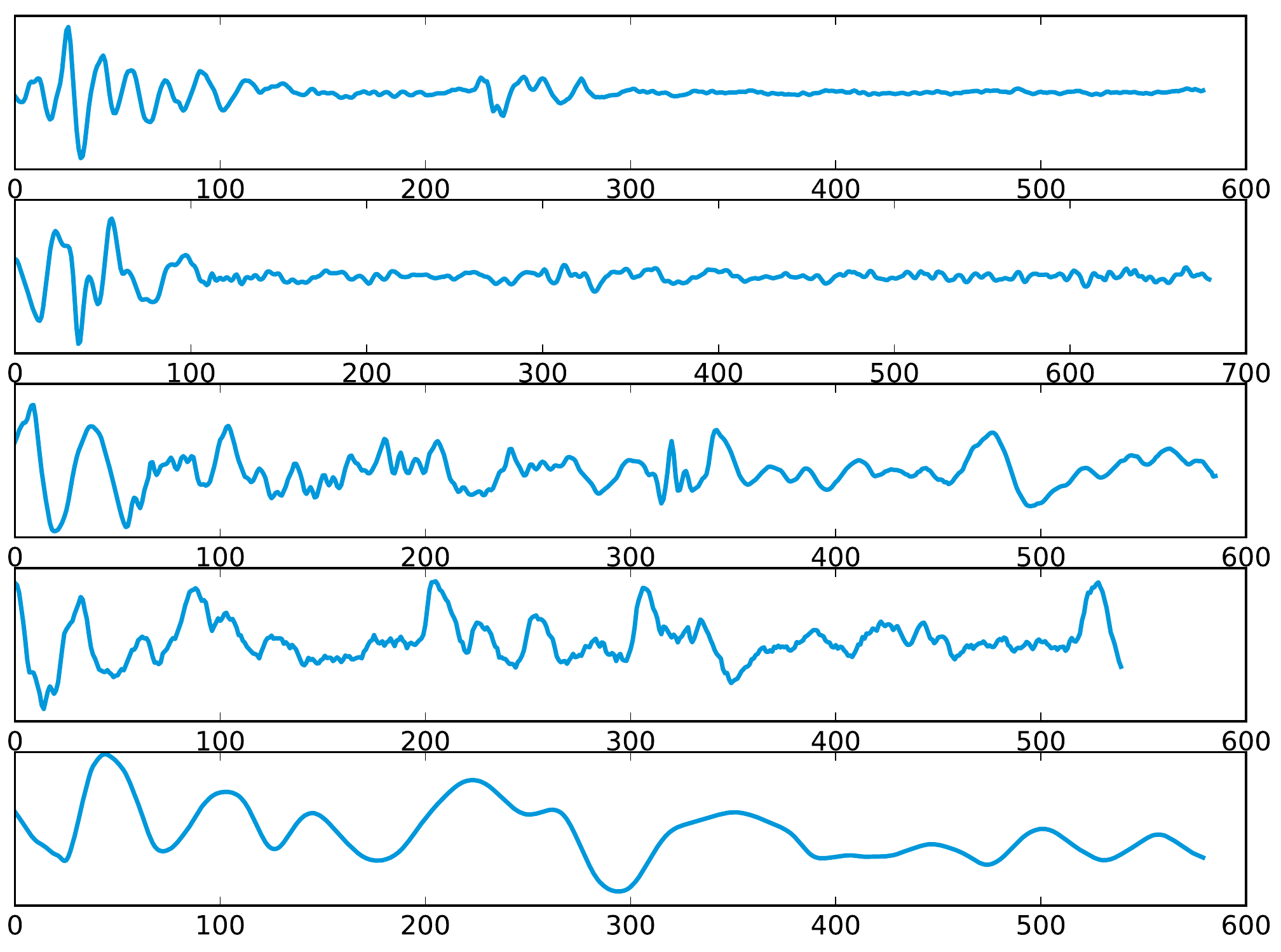} }
% \caption{}
\label{fig:segment}
\hspace{2em}
\subfigure[]{\includegraphics[width=0.49\textwidth]{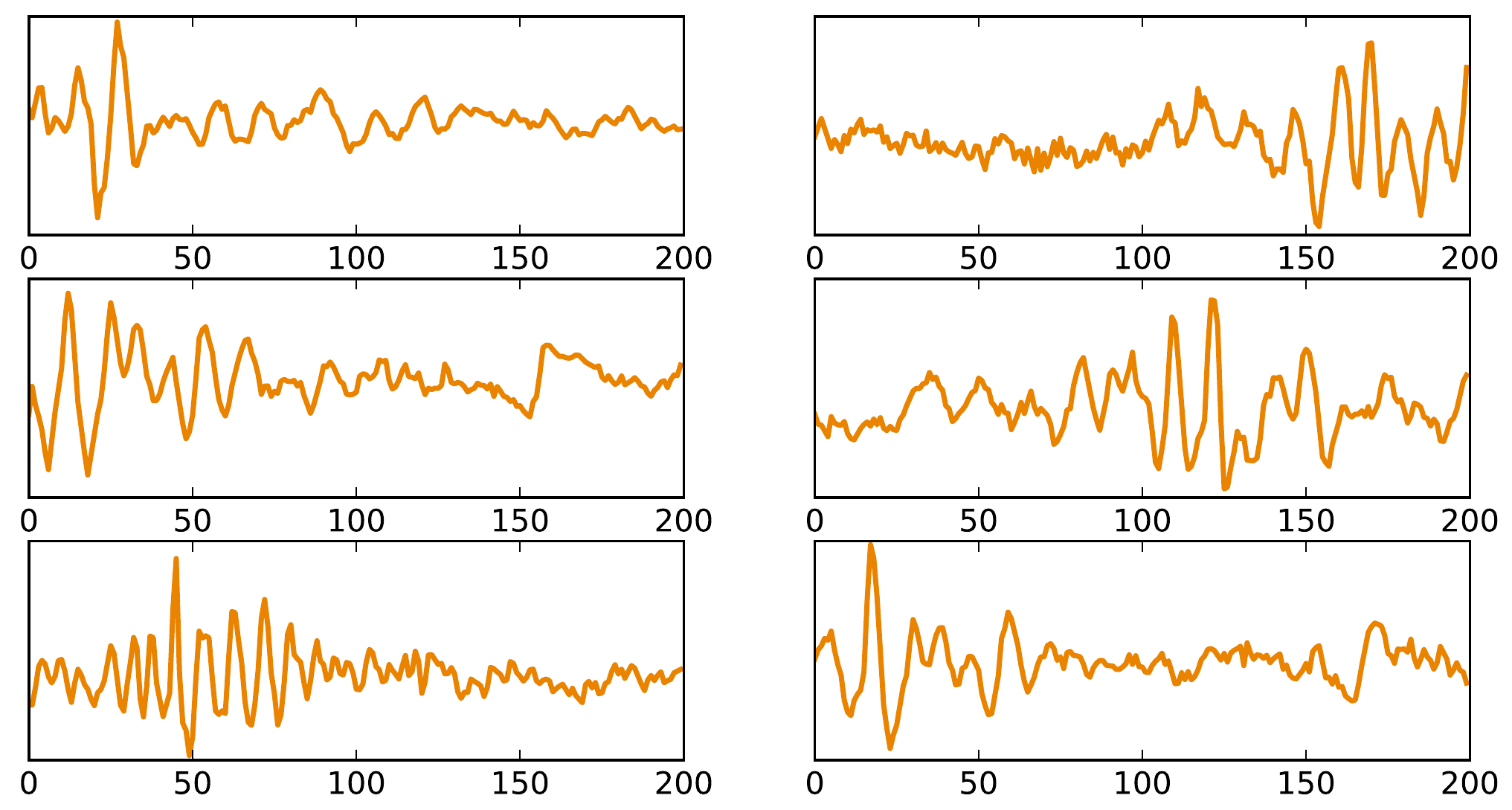}}
% \caption{}
\label{fig:fcnn-filter}
 
\caption{(a) Visualization of randomly chosen segments in the training data; (b) Visualization of learned filters (window size 200) by FCNN-Small model.}
\label{fig:vis-filters}
\end{figure}

To understand how the segmental convolutional neural networks work, we plot visualizations of randomly selected heart sound segments in the training dataset and filters learned by the FCNN-Small model in Fig.~\ref{fig:vis-filters}. The input segments have very different shapes, and noise is observable in some segments. This suggests the difficulty of the classification task. In addition, the visualization of filters shows that the network learned very good waveform-like patterns from the training data. This is even more convincing, considering the fact that all filters were randomly initialized prior to training. This qualitative result aligns very well with our intuitions about why CNN models are suitable for heart sound classification.

\begin{figure}[h]
\centering
\subfigure[]{
\includegraphics[width=0.4\textwidth]{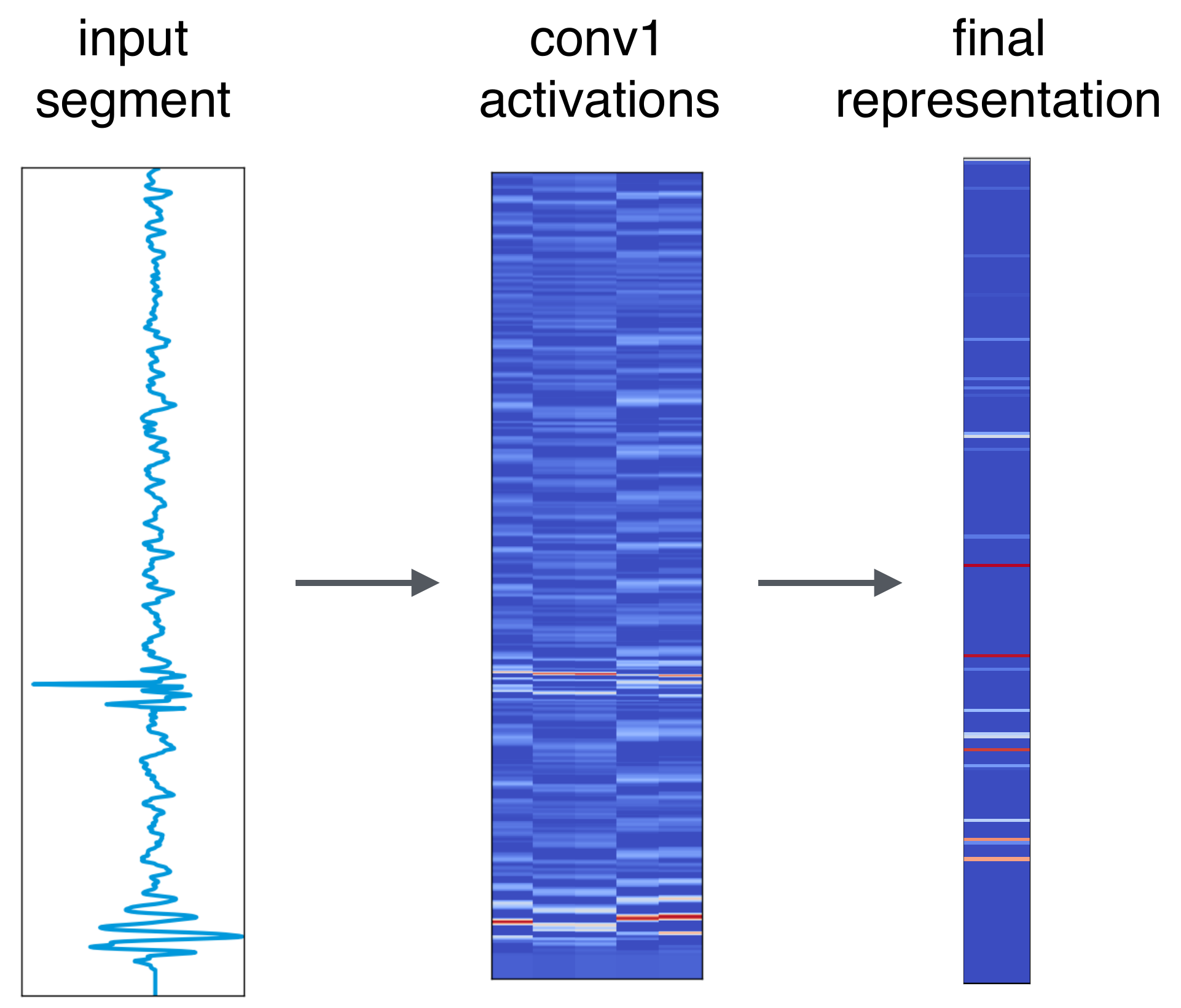}}
\label{fig:vis-normal}
\hspace{3em}
\subfigure[]{
\includegraphics[width=0.4\textwidth]{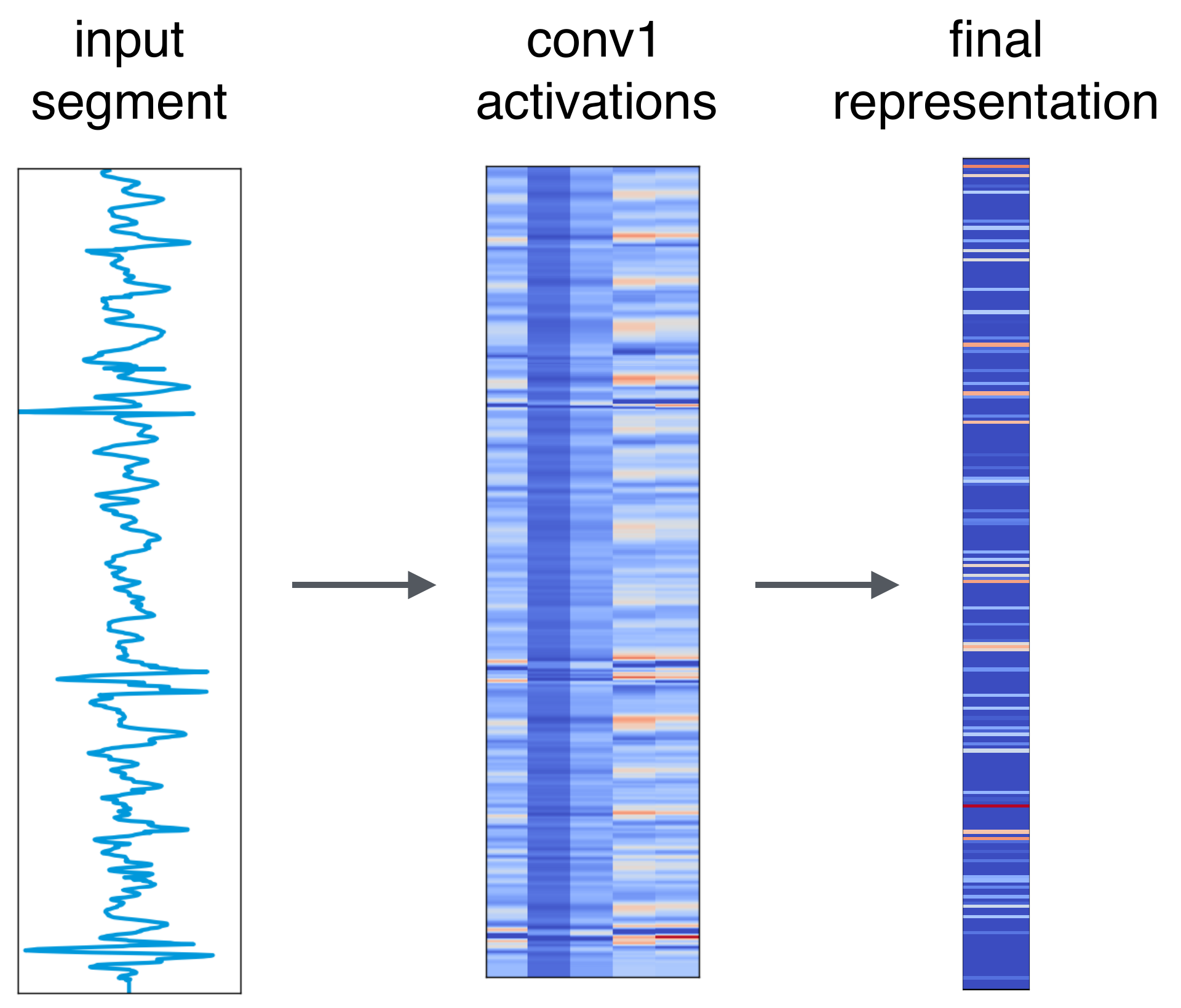}}
\label{fig:vis-abnormal}
 
\caption{Visualization of network activations for (a) a normal heart sound segment and (b) an abnormal segment. All activations are from DCNN-Deep model, and only activations in the first convolution layer and the final hidden layer are shown. Red color represents an activation value of 1, while blue color represents an activation value of 0.}
\label{fig:vis-activations}
\end{figure}

Fig.~\ref{fig:vis-activations} shows the network activations for normal and abnormal input segments. We find that, given the input segment, some of the output neurons in convolution layers activate, which indicates a pattern matched strongly with the signal at that local region, while others do not activate. Moreover, we find that more neurons in both the convolution layers and hidden layer are activated by abnormal segments compared to normal segments, indicating that many learned filters in the network are patterns of abnormal signals.

\subsection{Comparison}\label{sec4.4}

We compared the performance of the best performing CNN architectures, namely DCNN-Large and FCNN-Deep, to SVM, which was the best performing model among the traditional classifiers (Table~\ref{tab:cnn-result}). We see that CNNs outperform SVM significantly for accuracy and sensitivity. While FCNN-Deep has marginally better specificity, DCNN-Large has better performance in terms of accuracy and sensitivity. Our results show that application of CNNs to noisy heart sound recordings can produce better classification as compared to applying traditional classification techniques. Due to time limits, we were not able to fully explore the architecture space of the CNN models. Therefore, we believe that our segmental convolutional neural network approach has even more potential in classifying heart sound recordings than we have found.

\section{Discussion}\label{sec5}

Our investigation of the applicability of previously published work in traditional classification to noisy heart sound recordings suggests that further evaluation is needed. We found significant differences from feature extraction and classifier performance results reported from one such study, which justifies more rigorous scrutiny of previous work. Specifically, it would be useful to verify that feature extraction and traditional classification does indeed perform better on a dataset of clean heart sound recordings.

Due to limits of computing resources, we have not yet fully realized the potential of our CNN models. We believe that better-performing models with more filters and more layers can be achieved by doing a more thorough hyperparameter search. Another clear avenue of exploration is to decompose the signals further with EMD, which has been shown to delineate signals and noises of different origins in heart sound recordings\cite{Papadaniil:2014ki}. We would like to examine how splitting a recording into EMD components for use as separate input channels to our segmental CNNs may increase classification accuracy.

Limited by the annotation in the training data, our work is focused on the binary classification of heart sound recordings into normal and abnormal categories. However, it is also practically useful to predict a third ``unclassifiable'' category, especially when noise is dominant in the heard sound recordings. For example, in real world applications, this third label can serve as a signal for human intervention. Therefore another direction for future work is to explore the combination of supervised and unsupervised approaches to produce this ``unclassifiable'' label accurately.

\section{Conclusion}\label{sec6}

We propose a segmental convolutional neural network approach to accurately classify noisy heart sound recordings. We studies the effectiveness of two different types of convolutional neural network architectures, and compare their results with the application of traditional statistical classifiers on a set of manually curated features. Our results suggest that: First, traditional statistical classifiers using feature sets developed for low-noise recordings may perform worse on noisy recordings. Second, segmental convolutional neural networks with deep architectures and small filters can achieve higher accuracy in classifying noisy heart sound recordings without relying on manually-curated feature sets.

\section{Acknowledgements}

The authors would like to acknowledge Dr. Russ Altman, Dr. Steven Bagley and Dr. David Stark at Stanford University for their helpful suggestions to improve this work. We also want to thank Dr. Victor Froelicher for a helpful discussion on valvular heart diseases.

\bibliographystyle{ws-procs11x85}
\bibliography{main}

\begin{thebibliography}{10}

\bibitem{WHOGeneva2004}
W.~H. O.~E. Consultation, {\em Rheumatic Fever and Rheumatic Heart Disease},
  tech. rep., World Health Organization  (2001).

\bibitem{Carapetis16122008}
J.~R. Carapetis, {\em Circulation} {\bf 118}, 2748  (2008).

\bibitem{Marijon2008}
E.~Marijon, P.~Ou, D.~S. Celermajer, B.~Ferreira, A.~O. Mocumbi, D.~Sidi and
  X.~Jouven, {\em Bulletin of the World Health Organization} {\bf 86}, 84
  (2008).

\bibitem{Gersh2015}
B.~J. Gersh, Auscultation of cardiac murmurs in adults. In: UpToDate,  (2015).

\bibitem{Sztajzel2010308}
J.~M. Sztajzel, M.~Picard-Kossovsky, R.~Lerch, C.~Vuille and F.~P. Sarasin,
  {\em International journal of cardiology} {\bf 138}, 308  (2010).

\bibitem{Maglogiannis200947}
I.~Maglogiannis, E.~Loukis, E.~Zafiropoulos and A.~Stasis, {\em Computer
  methods and programs in biomedicine} {\bf 95}, 47  (2009).

\bibitem{lok1998accuracy}
C.~E. Lok, C.~D. Morgan and N.~Ranganathan, {\em CHEST Journal} {\bf 114}, 1283
   (1998).

\bibitem{Ishmail1987870}
A.~A. Ishmail, S.~Wing, J.~Ferguson, T.~A. Hutchinson, S.~Magder and K.~M.
  Flegel, {\em CHEST Journal} {\bf 91}, 870  (1987).

\bibitem{Jordan1987147}
M.~D. Jordan, C.~R. Taylor, A.~W. Nyhuis and M.~E. Tavel, {\em Archives of
  internal medicine} {\bf 147}, 721  (1987).

\bibitem{Vukanovic2006}
J.~M. Vukanovic-Criley, S.~Criley, C.~M. Warde, J.~R. Boker,
  L.~Guevara-Matheus, W.~H. Churchill, W.~P. Nelson and J.~M. Criley, {\em
  Archives of internal medicine} {\bf 166}, 610  (2006).

\bibitem{March20051443}
S.~K. March, J.~L. Bedynek and M.~A. Chizner, Teaching cardiac auscultation:
  effectiveness of a patient-centered teaching conference on improving cardiac
  auscultatory skills, in {\em Mayo Clinic Proceedings\/},  (11)2005.

\bibitem{Vukanovic2010}
J.~M. Vukanovic-Criley, A.~Hovanesyan, S.~R. Criley, T.~J. Ryan, G.~Plotnick,
  K.~Mankowitz, C.~R. Conti and J.~M. Criley, {\em Clinical cardiology} {\bf
  33}, 738  (2010).

\bibitem{Mangione1997}
S.~Mangione and L.~Z. Nieman, {\em Jama} {\bf 278}, 717  (1997).

\bibitem{Tavel15031996}
M.~E. Tavel, {\em Circulation} {\bf 93}, 1250  (1996).

\bibitem{Liang:1997tl}
H.~Liang, S.~Lukkarinen and I.~Hartimo, Heart sound segmentation algorithm
  based on heart sound envelogram, in {\em Computers in Cardiology 1997\/},
  1997.

\bibitem{Sun:2014eu}
S.~Sun, Z.~Jiang, H.~Wang and Y.~Fang, {\em Computer methods and programs in
  biomedicine} {\bf 114}, 219  (2014).

\bibitem{Oskiper:2002gh}
T.~Oskiper and R.~Watrous, Detection of the first heart sound using a
  time-delay neural network, in {\em Computers in Cardiology, 2002\/}, 2002.

\bibitem{Chen:2010va}
T.~Chen, K.~Kuan, L.~A. Celi and G.~D. Clifford, Intelligent heartsound
  diagnostics on a cellphone using a hands-free kit., in {\em AAAI Spring
  Symposium: Artificial Intelligence for Development\/}, 2010.

\bibitem{Schmidt:2010cy}
S.~Schmidt, C.~Holst-Hansen, C.~Graff, E.~Toft and J.~J. Struijk, {\em
  Physiological Measurement} {\bf 31}, p. 513  (2010).

\bibitem{Papadaniil:2014ki}
C.~D. Papadaniil and L.~J. Hadjileontiadis, {\em IEEE journal of biomedical and
  health informatics} {\bf 18}, 1138  (2014).

\bibitem{Leng:2015iy}
S.~Leng, R.~San~Tan, K.~T.~C. Chai, C.~Wang, D.~Ghista and L.~Zhong, {\em
  Biomedical engineering online} {\bf 14}, p.~1  (2015).

\bibitem{Uguz:2012fi}
H.~U{\u{g}}uz, {\em Journal of medical systems} {\bf 36}, 61  (2012).

\bibitem{Gharehbaghi:2015ga}
A.~Gharehbaghi, I.~Ekman, P.~Ask, E.~Nylander and B.~Janerot-Sjoberg, {\em
  International journal of cardiology} {\bf 198}, p.~58  (2015).

\bibitem{Saracoglu:2012wn}
R.~Sara{\c{c}}O{\u{g}}Lu, {\em Engineering Applications of Artificial
  Intelligence} {\bf 25}, 1523  (2012).

\bibitem{AvendanoValencia:2010ka}
L.~Avendano-Valencia, J.~Godino-Llorente, M.~Blanco-Velasco and
  G.~Castellanos-Dominguez, {\em Annals of Biomedical Engineering} {\bf 38},
  2716  (2010).

\bibitem{liu2016open}
C.~Liu, D.~Springer, Q.~Li, B.~Moody, R.~A. Juan, F.~J. Chorro, F.~Castells,
  J.~M. Roig, I.~Silva, A.~E. Johnson {\em et~al.}, {\em Physiological
  Measurement} {\bf 37}, p. 2181  (2016).

\bibitem{krizhevsky2012imagenet}
A.~Krizhevsky, I.~Sutskever and G.~E. Hinton, Imagenet classification with deep
  convolutional neural networks, in {\em Advances in neural information
  processing systems\/}, 2012.

\bibitem{simonyan2014very}
K.~Simonyan and A.~Zisserman, {\em arXiv preprint arXiv:1409.1556}   (2014).

\bibitem{wang2012end}
T.~Wang, D.~J. Wu, A.~Coates and A.~Y. Ng, End-to-end text recognition with
  convolutional neural networks, in {\em Pattern Recognition (ICPR), 2012 21st
  International Conference on\/}, 2012.

\bibitem{lecun1995convolutional}
Y.~LeCun and Y.~Bengio, {\em The handbook of brain theory and neural networks}
  {\bf 3361}, p. 1995  (1995).

\bibitem{Gradolewski:2014jma}
D.~Gradolewski and G.~Redlarski, {\em Computers in biology and medicine} {\bf
  52}, 119  (2014).

\bibitem{Singh2013}
M.~Singh and A.~Cheema, {\em International Journal of Computer Applications}
  {\bf 77}  (2013).

\bibitem{srivastava2014dropout}
N.~Srivastava, G.~Hinton, A.~Krizhevsky, I.~Sutskever and R.~Salakhutdinov,
  {\em The Journal of Machine Learning Research} {\bf 15}, 1929  (2014).

\bibitem{duchi2011adaptive}
J.~Duchi, E.~Hazan and Y.~Singer, {\em The Journal of Machine Learning
  Research} {\bf 12}, 2121  (2011).

\end{thebibliography}

\end{document}